\documentclass[a4paper]{aipproc}
\layoutstyle{6x9}
\title{Renormalization of HQET at Three Loops}

\author{A.~G.~Grozin}{
address={Budker Institute of Nuclear Physics, Novosibirsk 630090, Russia},
email={A.G.Grozin@inp.nsk.su},}

\begin{document}

\begin{abstract}
Three-loop propagator diagrams in HQET can be reduced,
using integration by parts, to 8 basis integrals:
5 trivial ones, two expressible via ${}_3\!F_2$,
and one only known up to $\varepsilon^0$.
Calculation of the heavy-quark propagator in HQET
is considerably simplified by the non-abelian exponentiation theorem.
\end{abstract}

\maketitle

\section{Three-loop propagator diagrams in HQET}

Heavy Quark Effective Theory (HQET) is an effective field theory
approximating QCD for problems with a single heavy quark
when characteristic momenta are much less than its mass $m$,
see the textbook~\cite{MW}.
A method of calculation of two-loop propagator diagrams in HQET
based on integration by parts~\cite{CT}
has been constructed in~\cite{BG}.
It has been extended to three loops in~\cite{G}.
Methods of perturbative calculations in HQET are reviewed in~\cite{Gr}.

\begin{figure}[ht]
\includegraphics{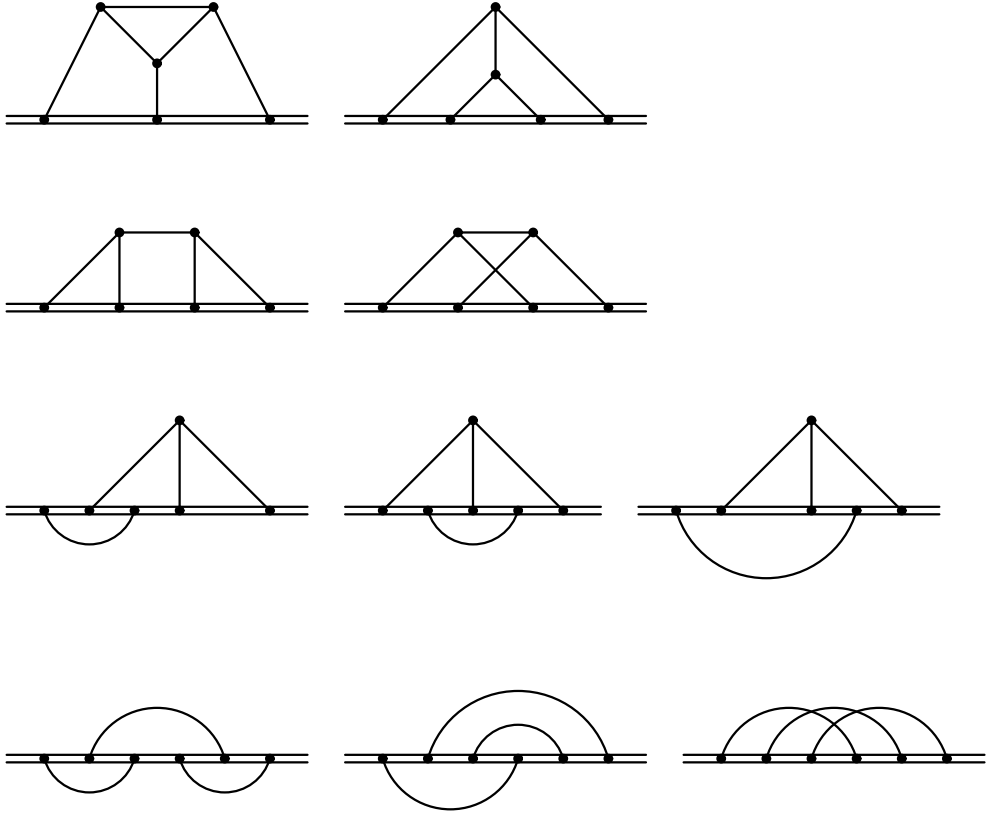}
\caption{Topologies of three-loop propagator diagrams in HQET}
\label{Top}
\end{figure}

\begin{figure}[ht]
\includegraphics{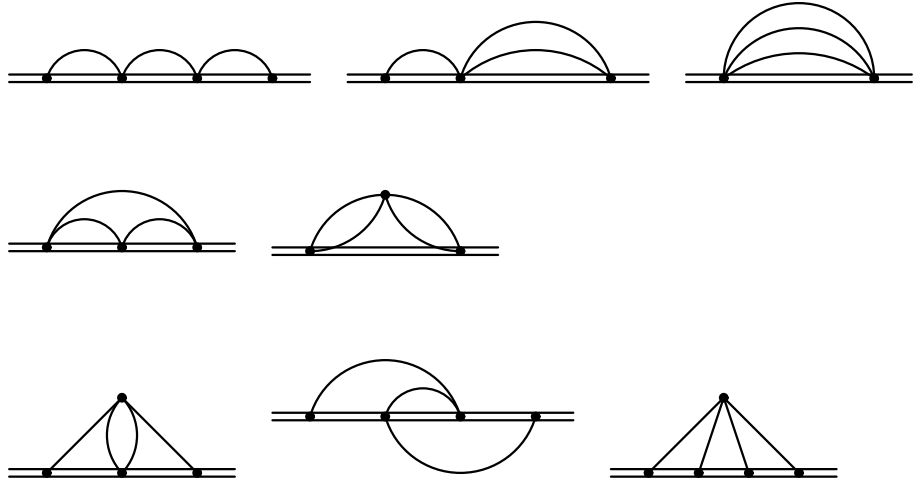}
\caption{Basis integrals}
\label{Bas}
\end{figure}

All generic topologies of three-loop propagator diagrams in HQET
are presented in Fig.~\ref{Top}.
The corresponding Feynman integrals depend on indices of the lines ---
the degrees of the corresponding denominators.
They can also contain powers of numerators which cannot be expressed
via the denominators.
If index of a line becomes zero, this line shrinks to a point.
In some cases, the resulting diagram contains lower-loop
propagator insertions, and is easy to calculate.
An algorithm which reduces any three-loop propagator integral in HQET
to a linear combination of 8 basis integrals
(with coefficients being rational functions of space-time dimension $d$)
has been constructed and implemented in~\cite{G}.
The basis integrals are presented in Fig.~\ref{Bas};
indices of all lines are equal to 1 here.

The first 5 basis integrals are trivial:
\begin{equation}
B_1 = I_1^3\,,\quad
B_2 = I_1 I_2\,,\quad
B_3 = I_3\,,\quad
B_4 = I_3 \frac{I_2}{I_1^2}\,,\quad
B_5 = I_3 \frac{G_2}{G_1^2}\,,
\label{B15}
\end{equation}
where
\begin{eqnarray}
&&I_n = \frac{\Gamma(1+2n\epsilon)\Gamma^n(1-\epsilon)}{(1-n(d-2))_{2n}}\,,
\nonumber\\
&&G_n =
\frac{1}{\left(n+1-n\frac{d}{2}\right)_n \left((n+1)\frac{d}{2}-2n-1\right)_n}\,
\frac{\Gamma(1+n\epsilon)\Gamma^{n+1}(1-\epsilon)}{\Gamma(1-(n+1)\epsilon)}
\label{sunset}
\end{eqnarray}
are the $n$-loop sunset HQET and massless integrals.

\begin{figure}[ht]
\begin{picture}(112,18)
\put(56,9){\makebox(0,0){\includegraphics{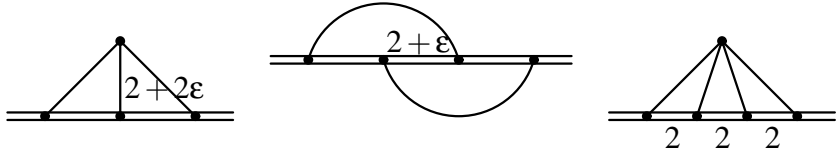}}}
\put(16.5,5){\makebox(0,0)[l]{$2+2\epsilon$}}
\put(55.5,9.5){\makebox(0,0)[b]{$2+\epsilon$}}
\put(89.3333,0){\makebox(0,0)[t]{2}}
\put(96,0){\makebox(0,0)[t]{2}}
\put(102.6667,0){\makebox(0,0)[t]{2}}
\end{picture}
\caption{Convergent integrals related to $B_{6\dots8}$}
\label{Three}
\end{figure}

The next two basis integrals are
\begin{equation}
B_6 = G_1 I(1,1,1,1,\epsilon)\,,\quad
B_7 = I_1 J(1,1,-1+2\epsilon,1,1)\,.
\label{B67}
\end{equation}
We use recurrence relations for the two-loop integrals $I$ and $J$
to re-express them via convergent integrals
(the first two diagrams in Fig.~\ref{Three})
$I(1,1,1,1,2+\epsilon)$ and $J(1,1,2+2\epsilon,1,1)$.
At a non-integer $n$, the integral $I(1,1,1,1,n)$ has been calculated in~\cite{BB},
and $J(1,1,n,1,1)$ --- in~\cite{G}.
In particular,
\begin{eqnarray}
&&\hspace{-6mm}I(1,1,1,1,2+\epsilon) =
\frac{8(3d-13)}{(d-4)(d-5)(d-6)(d-8)(2d-11)}
\frac{\Gamma(1+6\epsilon)\Gamma(1-2\epsilon)\Gamma(1-\epsilon)}{\Gamma(1+\epsilon)}
\nonumber\\
&&{}\times\Biggl[ {}_3\!F_2 \biggl(
\begin{array}{c} 1, 2-2\epsilon, 3+4\epsilon\\ 3+\epsilon, 4+4\epsilon \end{array}
\biggr.\biggl|\, 1 \biggr)
\nonumber\\
&&\hspace{6mm} - \frac{(d-6)(d-8)(2d-11)}{12(d-3)(d-4)(3d-13)}
\frac{\Gamma^2(1+3\epsilon)\Gamma(1-3\epsilon)\Gamma(1+\epsilon)}
{\Gamma(1+6\epsilon)\Gamma(1-2\epsilon)}
\Biggr]\,,
\nonumber\\
&&\hspace{-6mm}J(1,1,2+2\epsilon,1,1) =
\frac{1}{3(d-4)(d-5)(d-6)(2d-9)}
\frac{\Gamma(1+6\epsilon)\Gamma^2(1-\epsilon)}{\Gamma(1+2\epsilon)}
\nonumber\\
&&{}\times{}_3\!F_2 \biggl(
\begin{array}{c} 1, 2-2\epsilon, 1+4\epsilon\\ 3+2\epsilon, 2+4\epsilon \end{array}
\biggr.\biggl|\, 1 \biggr)\,.
\label{F6}
\end{eqnarray}
Using the methods of~\cite{BBB}, it is not difficult to obtain the expansions
\begin{eqnarray}
&&{}_3\!F_2 \biggl(
\begin{array}{c} 1, 2-2\epsilon, 3+4\epsilon\\ 3+\epsilon, 4+4\epsilon \end{array}
\biggr.\biggl|\, 1 \biggr) =
3 + \left(18\zeta_2-\frac{73}{2}\right) \epsilon + 3(-48\zeta_3-7\zeta_2+75) \epsilon^2
\nonumber\\
&&\quad{} + 3(216\zeta_4+56\zeta_3+115\zeta_2-506) \epsilon^3
\nonumber\\
&&\quad{}
+ 12(-375\zeta_5+90\zeta_2\zeta_3-63\zeta_4-230\zeta_3-135\zeta_2+789) \epsilon^4
+ \cdots
\nonumber\\
&&{}_3\!F_2 \biggl(
\begin{array}{c} 1, 2-2\epsilon, 1+4\epsilon\\ 3+2\epsilon, 2+4\epsilon \end{array}
\biggr.\biggl|\, 1 \biggr) =
2 + 6(-2\zeta_2+3) \epsilon + 12(10\zeta_3-11\zeta_2+6) \epsilon^2
\nonumber\\
&&\quad{} + 24(-28\zeta_4+55\zeta_3-27\zeta_2+9) \epsilon^3
\nonumber\\
&&\quad{}
+ 48(94\zeta_5-16\zeta_2\zeta_3-154\zeta_4+135\zeta_3-45\zeta_2+12) \epsilon^4 + \cdots
\label{Fexp}
\end{eqnarray}
They can be extended to $\epsilon^8$, should it be needed.

The most difficult basis integral $B_8=J_c(1,1,1,1,1,1,1)$
can be re-expressed via a convergent integral $J_c(2,2,2,1,1,1,1)$
(the last diagram in Fig.~\ref{Three}).
It can be calculated in 4-dimensional euclidean coordinate space
in cylindrical coordinates.
Taking first the integral in the radial coordinate of the light vertex,
then the integral in its time, and then the integrals in the two
times of heavy-light vertices, we obtain
\begin{equation}
J_c(2,2,2,1,1,1,1) = 2(\zeta_3-1) + \mathcal{O}(\epsilon)\,.
\label{B8}
\end{equation}

\section{Heavy-quark propagator in HQET}

Due to the non-abelian exponentiation theorem~\cite{GFT},
the unrenormalized HQET quark propagator in the coordinate space can be written as
\begin{eqnarray}
&&\hspace{-5mm}S_0(t) = -i\theta(t) \exp \Biggl[
C_F \frac{g_0^2}{(4\pi)^{d/2}} \left(\frac{it}{2}\right)^{2\epsilon} s
+ C_F \frac{g_0^4}{(4\pi)^d} \left(\frac{it}{2}\right)^{4\epsilon}
\left(C_A s_A + T_F n_l s_l\right)
\nonumber\\
&&\hspace{-5mm}\quad{}
+ C_F \frac{g_0^6}{(4\pi)^{3d/2}} \left(\frac{it}{2}\right)^{6\epsilon}
\left(C_A^2 s_{AA} + C_F T_F n_l s_{lF} + C_A T_F n_l s_{lA}
+ \left(T_F n_l\right)^2 s_{ll} \right) + \cdots \Biggr]\,.
\label{S0}
\end{eqnarray}

\begin{figure}[ht]
\begin{picture}(107,38)
\put(53.5,19.5){\makebox(0,0){\includegraphics{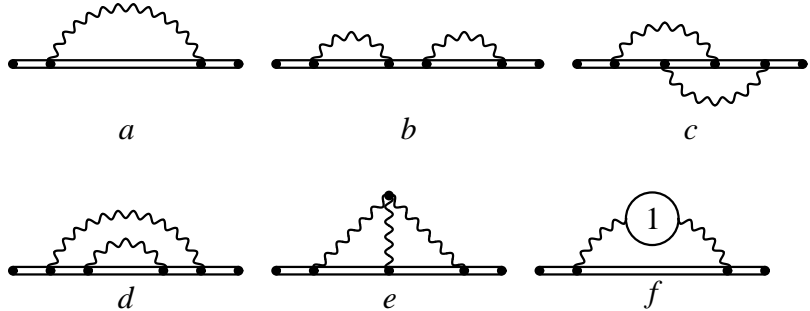}}}
\put(16,19.5){\makebox(0,0)[b]{$a$}}
\put(53.5,19.5){\makebox(0,0)[b]{$b$}}
\put(91,19.5){\makebox(0,0)[b]{$c$}}
\put(16,-3){\makebox(0,0)[b]{$d$}}
\put(51,-3){\makebox(0,0)[b]{$e$}}
\put(86,-3){\makebox(0,0)[b]{$f$}}
\put(86,9){\makebox(0,0){1}}
\end{picture}
\caption{One- and two-loop diagrams for the HQET propagator}
\label{Prop12}
\end{figure}

\begin{figure}[ht]
\begin{picture}(142.2,54.9)
\put(71.1,27.45){\makebox(0,0){\includegraphics[scale=0.9]{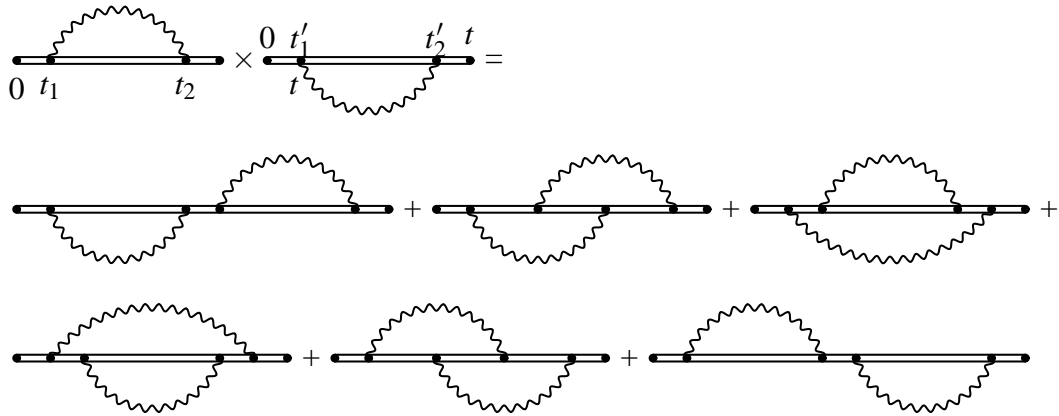}}}
\put(0.9,45){\makebox(0,0)[t]{0}}
\put(5.4,45){\makebox(0,0)[t]{$t_1$}}
\put(23.4,45){\makebox(0,0)[t]{$t_2$}}
\put(37.9,45){\makebox(0,0)[t]{$t$}}
\put(34.2,51.75){\makebox(0,0)[t]{0}}
\put(38.7,52.2){\makebox(0,0)[t]{$t_1'$}}
\put(56.7,52.2){\makebox(0,0)[t]{$t_2'$}}
\put(61.2,51.75){\makebox(0,0)[t]{$t$}}
\put(31.05,47.25){\makebox(0,0){$\times$}}
\put(64.35,47.25){\makebox(0,0){=}}
\put(53.55,27.45){\makebox(0,0){+}}
\put(95.85,27.45){\makebox(0,0){+}}
\put(138.15,27.45){\makebox(0,0){+}}
\put(40.05,7.65){\makebox(0,0){+}}
\put(82.35,7.65){\makebox(0,0){+}}
\end{picture}
\caption{Exponentiation theorem}
\label{Exp}
\end{figure}

Suppose we have calculated the one-loop correction to the HQET propagator
in the coordinate space (Fig.~\ref{Prop12}$a$).
Let's multiply this correction by itself (Fig.~\ref{Exp}).
We get an integral in $t_1$, $t_2$, $t_1'$, $t_2'$
with $0<t_1<t_2<t$, $0<t_1'<t_2'<t$.
Ordering of primed and non-primed integration times can be arbitrary.
The integration area is subdivided into 6 regions,
corresponding to 6 diagrams in Fig.~\ref{Exp}.
If the colour factors of the diagrams Fig.~\ref{Prop12}$c$, $d$
were the same as that of the one-particle-reducible diagram (Fig.~\ref{Prop12}$b$),
i.~e.\ equal to the square of the colour factor $C_F$
of the one-loop diagram (Fig.~\ref{Prop12}$a$) (as in the abelian case),
then the sum of the diagrams Fig.~\ref{Prop12}$b$, $c$, $d$ would be equal
to $\frac{1}{2}$ of the square of the one-loop correction (Fig.~\ref{Prop12}$a$),
as given by the square of the first term in the expansion of the exponent~(\ref{S0}).
In the non-abelian case, however, the colour factor of Fig.~\ref{Prop12}$c$
differs from $C_F^2$ by $-C_F C_A/2$,
which is the colour factor of Fig.~\ref{Prop12}$e$.
This is because when we reduce the colour factor of Fig.~\ref{Prop12}$c$
to that of the reducible diagram Fig.~\ref{Prop12}$b$,
we get an extra term with the commutator $[t^a,t^b]$,
which has the colour structure of Fig.~\ref{Prop12}$e$.
Therefore, we should include the contribution of Fig.~\ref{Prop12}$c$
with $-C_F C_A/2$ instead of its full colour factor into the term $s_A$
in the exponent~(\ref{S0}).
This part of the colour factor is called maximally non-abelian
or colour-connected~\cite{GFT}.
Of course, the diagram Fig.~\ref{Prop12}$e$ also contributes to $s_A$;
the diagram Fig.~\ref{Prop12}$f$, with the one-loop gluon self-energy correction,
contributes to $s_l$ (quark loop) and $s_A$ (gluon and ghost loops).

\begin{figure}[ht]
\begin{picture}(141.3,101.7)
\put(70.65,51.3){\makebox(0,0){\includegraphics[scale=0.9]{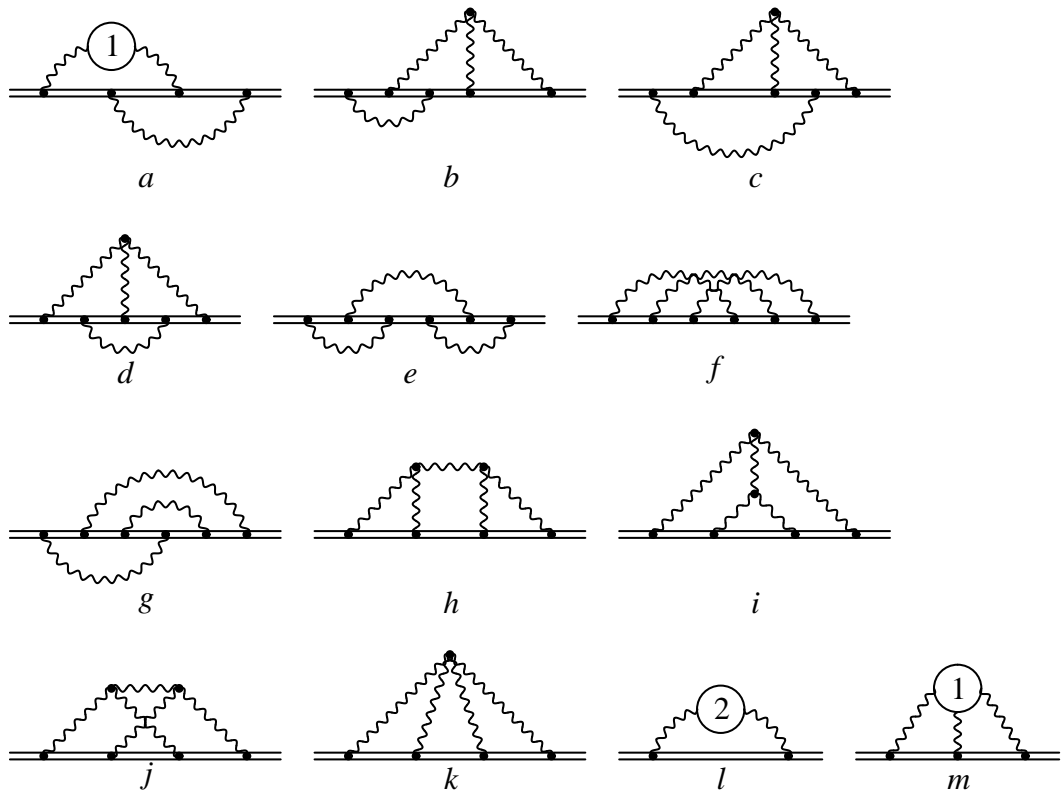}}}
\put(18.9,77.4){\makebox(0,0)[b]{$a$}}
\put(59.4,77.4){\makebox(0,0)[b]{$b$}}
\put(99.9,77.4){\makebox(0,0)[b]{$c$}}
\put(16.2,51.3){\makebox(0,0)[b]{$d$}}
\put(54,51.3){\makebox(0,0)[b]{$e$}}
\put(94.5,51.3){\makebox(0,0)[b]{$f$}}
\put(18.9,20.7){\makebox(0,0)[b]{$g$}}
\put(59.4,20.7){\makebox(0,0)[b]{$h$}}
\put(99.9,20.7){\makebox(0,0)[b]{$i$}}
\put(18.9,-2.7){\makebox(0,0)[b]{$j$}}
\put(59.4,-2.7){\makebox(0,0)[b]{$k$}}
\put(95.4,-2.7){\makebox(0,0)[b]{$l$}}
\put(126.9,-2.7){\makebox(0,0)[b]{$m$}}
\put(14.4,96.3){\makebox(0,0){1}}
\put(95.4,8.1){\makebox(0,0){2}}
\put(126.9,10.8){\makebox(0,0){1}}
\end{picture}
\caption{Three-loop diagrams contributing to the three-loop term
in the exponent~(\ref{S0})}
\label{Prop3}
\end{figure}

In the same way, we can consider the set of three-loop diagrams
obtained by multiplying the corrections of Fig.~\ref{Prop12}$a$ and $f$.
We can imagine that this set is obtained from the one-particle-reducible diagram
by allowing the gluon -- heavy-quark vertices to slide along the heavy-quark line,
crossing each other.
These diagrams are said to contain two connected webs (or c-webs)~\cite{GFT}.
Everything is already accounted for by the product
of the one-loop correction and the (Fig.~\ref{Prop12}$f$ part of) two-loop correction
in the expansion of the exponent~(\ref{S0}),
except the contribution of Fig.~\ref{Prop3}$a$ (and its mirror-symmetric),
taken with the maximally non-abelian part of its colour factor.
It contributes to the three-loop correction in the exponent.
Similarly, out of all the diagrams with two connected webs
Fig.~\ref{Prop12}$a$ and $e$,
only those of Fig.~\ref{Prop3}$b$, $c$ (plus their mirror-symmetric ones), and $d$
contribute to $s_{AA}$, with the maximally non-abelian part of their colour factors.
This part appears, in the case of Fig.~\ref{Prop3}$b$, for example,
when we commute $t^a$ matrices to obtain the colour structure of the reducible diagram;
it is identical to the colour factor of Fig.~\ref{Prop3}$h$,
equal to $C_F C_A^2/4$.

Now we are going to consider all the diagrams with three gluons,
both ends of which are attached to the heavy-quark line.
They are said to contain three c-webs,
each of them is that of Fig.~\ref{Prop12}$a$.
We decompose the colour factors of these 15 diagrams in the following way.
We move the vertices along the heavy-quark line in such a way
as to disentangle those c-webs.
While doing so, we get extra terms from the commutators,
having colour structures of the corresponding diagrams with the three-gluon vertex.
These diagrams have fewer c-webs, which are more complicated.
Finally, each colour factor can be expressed as a linear combination of three ones:
$C_F^3$ (3 c-webs of Fig.~\ref{Prop12}$a$);
$-C_F^2 C_A/2$ (2 c-webs of Fig.~\ref{Prop12}$a$ and $e$);
$C_F C_A^2/4$ (1 c-web of Fig.~\ref{Prop3}$h$).
The first one occurs with the unit coefficient in all 15 colour factors.
The sum of the corresponding contributions is just the term with the cube
of the one-loop correction in the expansion of the exponent~(\ref{S0}).
The second colour structure occurs in the diagrams
obtained by multiplying Fig.~\ref{Prop12}$a$ and $c$;
the sum of the corresponding contributions is contained in the product
of the one-loop term and the two-loop one in the expansion of the exponent.
We are left with the colour-connected parts of the colour factors
(a single c-web contributions).
They are present in the diagrams Fig.~\ref{Prop3}$e$, $f$, $g$
(and its mirror-symmetric).
They contribute to $s_{AA}$ in~(\ref{S0}).

\begin{figure}[ht]
\begin{picture}(92,22)
\put(46,11){\makebox(0,0){\includegraphics{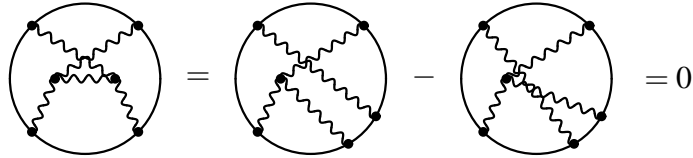}}}
\put(26,11){\makebox(0,0){${}={}$}}
\put(56,11){\makebox(0,0){${}-{}$}}
\put(88,11){\makebox(0,0){${}=0$}}
\end{picture}
\caption{Vanishing colour factor}
\label{Col0}
\end{figure}

We are left with the diagrams containing only a single connected web.
Those of Fig.~\ref{Prop3}$h$ and $i$ have equal colour factors
(this is evident if we close the quark line),
they contribute to $s_{AA}$.
The colour factor of Fig.~\ref{Prop3}$j$ is zero.
This becomes clear if we close the quark line
and write a three-gluon vertex as the commutator (Fig.~\ref{Col0}).
The diagram with the four-gluon vertex (Fig.~\ref{Prop3}$k$)
can be decomposed into three terms,
with colour factors of Fig.~\ref{Prop3}$h$, $i$, $j$.
The diagram Fig.~\ref{Prop3}$l$ contains two-loop gluon self-energy corrections,
including one-particle-reducible ones;
it contributes to $s_{AA}$, $s_{lF}$, $s_{lA}$, $s_{ll}$.
The diagram Fig.~\ref{Prop3}$m$ contains one-loop corrections
to the three-gluon vertex, including one-particle-reducible ones
(i.~e., one-loop self-energy corrections to each gluon propagator
in Fig.~\ref{Prop12}$e$);
it contributes to $s_{AA}$, $s_{lA}$.

In order to find the heavy-quark field renormalization constant,
we should re-express the unrenormalized propagator~(\ref{S0})
in terms of the renormalized coupling $\alpha_s(\mu)$ and
gauge-fixing parameter $a(\mu)$ instead of $g_0^2$ and $a_0$,
and require that the renormalized propagator
$S(t)=Z_Q^{-1}S_0(t)$ is finite in the limit $\epsilon\to0$.
After this re-expression, $S_0(t)$ still has the exponential form
with the same colour structures as in~(\ref{S0}).
Therefore, the renormalization constant can be written in the exponential form, too:
\begin{eqnarray}
&&Z_Q = \exp \Biggl[ C_F \frac{\alpha_s}{4\pi} z
+ C_F \left(\frac{\alpha_s}{4\pi}\right)^2 \left(C_A z_A + T_F n_l z_l\right)
\nonumber\\
&&\qquad{} + C_F \left(\frac{\alpha_s}{4\pi}\right)^3
\left(C_A^2 z_{AA} + C_F T_F n_l z_{lF} + C_A T_F n_l z_{lA}
+ \left(T_F n_l\right)^2 z_{ll}\right) + \cdots \Biggr]\,.
\label{ZQ}
\end{eqnarray}
The coefficients $z$ are obtained simply by singling out poles in $\epsilon$
from the corresponding terms in the exponent for $S_0(t)$
(expressed via the renormalized quantities).

The heavy-quark field anomalous dimension in HQET is
defined as $\gamma_Q=d\log Z_Q/d\log\mu$.
The $\mu$-dependence comes from $\alpha_s(\mu)$ and $a(\mu)$ in $Z_Q$.
Using $\beta(\alpha_s)$ and $\gamma_A(\alpha_s)$
for their differentiation, we obtain
\begin{eqnarray}
&&\hspace{-6mm}
\gamma_Q = 2(a-3) C_F \frac{\alpha_s}{4\pi}
+ \left[\left(\frac{a^2}{2}+4a-\frac{179}{6}\right) C_A + \frac{32}{3} T_F n_l \right]
C_F \left(\frac{\alpha_s}{4\pi}\right)^2
\label{gammaQ}\\
&&\hspace{-6mm} + \Biggl[
\biggl(\frac{5}{8}a^3 + \frac{3}{4}\left(\zeta_3+\frac{13}{4}\right)a^2
+ \left(-8\zeta_4+6\zeta_3+\frac{271}{16}\right)a
-24\zeta_4-\frac{123}{4}\zeta_3-\frac{23815}{216}\biggr) C_A^2
\nonumber\\
&&\hspace{-6mm} - 6(16\zeta_3-17) C_F T_F n_l
+ \left(-\frac{17}{2}a+96\zeta_3+\frac{782}{27}\right) C_A T_F n_l
+ \frac{160}{27} \left(T_F n_l\right)^2
\Biggr] C_F \left(\frac{\alpha_s}{4\pi}\right)^3\,.
\nonumber
\end{eqnarray}
The two-loop term contains no $C_F^2$ part,
and the three-loop one contains no $C_F^3$ and $C_F^2 C_A$ parts,
as a consequence of the non-abelian exponentiation theorem.
The two-loop term was obtained in~\cite{BG}.
The three-loop term has been obtained,
by a completely different method, in~\cite{MR};
our calculation confirms this result.

I am grateful to D.~J.~Broadhurst, K.~G.~Chetyrkin and T.~van~Ritbergen
for useful discussions.

\end{document}